# Publish for Public: Improving Access of Public Libraries' Users to Research Findings through Plain Language Summaries

**Behrooz Rasuli**

Iranian Research Institute for Information Science and Technology (IranDoc)

**Abstract**

Public libraries play a crucial role in disseminating knowledge to society. However, most of their users do not have specialized knowledge to understand the new research findings. Providing plain language summaries (PLSs) in public libraries is a way to make the new research findings more accessible and understandable for the public. This article proposes a framework for providing PLSs as a new service in public libraries. Drawing from literature on science and society, PLSs, and public libraries, a theoretical framework is developed. The findings suggest that public libraries can collect PLSs through different methods, such as professional teams, researchers, crowdsourcing, etc. Library newsletters, special publications, brochures, independent online databases, and social networks are among the most effective for making PLSs accessible for the users. By proposing a framework for providing PLSs in public libraries, this study helps to bridge the gap between scientific research and the public.

**Introduction**

Public libraries have long been recognized as important information centers, uniquely positioned to meet the diverse needs of society and address its issues (Choy 2007). They serve as valuable institutions that understand and respond to the community's needs due to their direct and continuous connection with general users. No other organization can assume the role of public libraries in different societies.

In the 20th century, public libraries assumed a greater responsibility beyond being mere repositories of information sources. They transformed into "universities of the public," providing access to research and scientific findings for users of all backgrounds, from the illiterate to experts in various fields (Kent 2002). This broader role is commonly known today as the "promotion of science" (Vrana 2010).

Promoting science is an integral part of scholarly communication, offering opportunities for the wider society to gain understanding and insights from science and scientific findings (Martínez Silvagnoli et al. 2022). The link between science and people's everyday lives necessitates their engagement in monitoring scientific progress, as it directly affects their quality of life (Shahriari and Rasuli 2021). With the increasing emphasis on research impact during the last decade, the importance of science promotion has grown significantly. Research impact refers to the influence and effect that research outcomes have on various aspects of society, including economy, culture, environment, health, etc. It assesses the significance and value of research beyond traditional academic metrics such as publications and citations.

Public libraries, being uniquely connected to the public and society, are considered ideal institutions for promoting science (Choy 2007). In recent decades, public libraries have sought to strengthen their partnerships with universities and scientific centers to enhance their science promotion services (Borgman 2003). The role of science promotion in public libraries is recognized worldwide (Mumelaš, Martek, and Mučnjak 2022).

Among the different methods of science promotion, one that has gained attention in recent years is the publication of scientific findings in plain language. Scientific research often employs technical terminology that is challenging for the general public to grasp. Translating complex terms into simpler, universally understandable language can facilitate the advancement of science. Consequently, researchers are encouraged to provide concise and comprehensive summaries of their work alongside their research reports, known as "lay summaries" or "plain language summary" (Lobban, Gardner, and Matheis 2022; Smith 2009).

While public libraries have made efforts to promote science through conventional methods employed by other institutions, they may not fully recognize the potential of plain language summaries (PLSs). This article aims to develop a process and mechanism for bridging science and society through the provision of PLSs as a new service in public libraries. The article discusses the definition of plain language summary, methods for collection, potential distribution channels, and support mechanisms for users of summaries.

**Providing PLSs as a Library Service**

The Public may not completely understand the content of a scientific work because these works are written in specialized language and are filled with complex and technical terms that perhaps only researchers in a specific scientific field can understand. For example, a farmer who is engaged in the profession of farming may have a proper understanding of the practical terminology in the field of agriculture, but (s)he might struggle to understand the specialized and technical terms in the field of genetic engineering. In such situations, "plain language summaries" can bridge the gap between researchers and the general public, allowing them to read each other's language more easily.

PLSs are concise summaries that accompany scholarly journal articles, scientific reports, etc. written in a clear and jargon-free style to ensure easy accessibility and comprehension. These summaries aim to accurately convey the scientific messages and conclusions presented in the original publication. Typically, PLSs undergo peer review as part of the manuscript submission process. While PLSs are a relatively recent addition to scholarly publishing, they have long been a prominent feature of research outputs produced by institutions like Cochrane and the National Institute for Health and Care Research in the United Kingdom (Rosenberg et al. 2023). A piece of communication is considered to be in plain language summary when its content, organization, and presentation are designed in a clear and accessible manner, allowing the intended readers to effortlessly locate the information they require, comprehend what they discover, and effectively utilize that information (Gainey et al. 2023).

Many believe that PLSs can enhance the impact of scientific works and findings (Sedgwick et al. 2021). In addition to increasing impact, the publication of PLSs aims to make research findings more accessible to everyone so that they can (perhaps) benefit from scientific research findings. There is also an ethical consideration. In other words, research ethics implies that researchers present their research findings in a way that encompasses a broader range of readers (Shahriari and Rasuli 2021). These summaries are very important in science communication and some key organizations, such as the "European Union" has recommended the writing of such summaries for medical research in its regulations (European Commission 2018). Even some scientific journals today require the inclusion of PLSs for research articles (FitzGibbon et al. 2020).

The study by Shahriari and Rasuli (2021) demonstrated that PLSs could lead to better and more efficient understanding of research findings by the general public. In this study, PLSs and technical abstracts were provided to the public, and their understanding of the summaries and abstracts was assessed through questions. The results showed a statistically significant difference in the understanding of PLSs compared to technical abstracts, indicating that people connect more easily with PLSs.

Considering the nature and function of PLSs, the users of these summaries encompass a wide range of people. Since public library users are also part of this diverse range, the provision of these summaries in public libraries can be proposed as a recommended service. Thus, providing and presenting these summaries to library users is an important service that may target specific users. For example, main users of agricultural research are farmers who are engaged in this profession. Many of these farmers may not have university education or live in rural areas where access to scientific centers is difficult. Therefore, public libraries, by developing this service, can meet many of these users' information needs. Moreover, this special service will also benefit ordinary users. As a result, public libraries should design a specific mechanism for providing this service. The following section discusses the most important methods that can be used to provide PLSs in public libraries.

The provision of PLSs in public library contexts aligns with existing initiatives to make scholarly research more accessible. Several initiatives have been undertaken in the library field to bridge the gap between scholarly research and the general public, and the inclusion of PLSs contributes to these efforts. One existing initiative is the promotion of open access resources in public libraries (Scott 2011). By incorporating PLSs into their open access collections, public libraries enhance the accessibility of research for individuals who may not have the expertise or background to engage with traditional scholarly publications. PLSs provide a concise and user-friendly overview of research findings, facilitating easier understanding and engagement.

Another initiative is the development of research literacy programs in public libraries (Hall 2010). These programs aim to equip library users with the skills necessary to navigate scholarly research effectively. By incorporating PLSs into these programs, libraries can enhance the learning experience by providing clear and simplified explanations of complex research topics. PLSs serve as a valuable resource for individuals seeking to improve their research literacy, enabling them to engage with scholarly content with greater confidence.

Furthermore, the integration of PLSs aligns with the movement towards inclusive and user-centered library services. Public libraries strive to meet the diverse information needs of their communities, including individuals with varying levels of expertise and backgrounds. By providing PLSs, libraries ensure that research is presented in a format that is accessible and understandable to a broader audience. This initiative promotes inclusivity by breaking down barriers to accessing scholarly information and empowering individuals who may feel intimidated by the technical language often found in academic publications. Additionally, some public libraries have established partnerships with academic institutions and researchers to enhance access to scholarly research (Wynia Baluk et al. 2023). These collaborations often involve the provision of research databases or access to academic journals. By including PLSs alongside these resources, libraries enhance the value of these partnerships by facilitating easier comprehension and engagement with the research content. In summary, the inclusion of PLSs in public library contexts overlaps with existing initiatives to make scholarly research more accessible. By aligning with open access initiatives, research literacy programs, user-centered services, and collaborative partnerships, libraries can further enhance the accessibility and usability of scholarly research for their patrons. The integration of PLSs serves as a valuable contribution to these initiatives, supporting the goal of making research more inclusive, understandable, and relevant to the general public.

**Methods for Providing PLSs in Public Libraries**

PLSs can be prepared in various ways in public libraries and made available to library users. The first method is to request researchers and authors themselves to prepare PLSs for their research findings published in the form of articles or other forms. To do this, librarians must constantly monitor scientific and academic research and, if they come across suitable research aligned with the library's mission and the needs of users, contact to the authors of these works and ask them to prepare PLSs for their works. In such circumstances, it is necessary to train the authors and researchers for writing PLSs, and librarians should provide them with the necessary training and relevant guidelines (for finding current guideline to prepare PLSs, see [Appendix 1](#)). Although the best persons for writing PLSs are researchers who have conducted and published specific study, they may not have the required skills to write such summaries (Shahriari and Rasuli 2021).

One of the main challenges of this approach is that authors and researchers of scientific works may not be motivated to collaborate with the library in preparing PLSs due to the limited time (Kuehne and Olden 2015). Therefore, this service may not be transformed into a systematic process through the establishing this method. Nevertheless, the costs for libraries to provide the service of offering PLSs to library users will be negligible. Another benefit of this approach is that public libraries and librarians choose which works are suitable for summarization. Naturally, this selection will be based on the library's needs and the demands of the community of users. In any case, if the library chooses this method to provide PLSs, it should consider the necessary motivations and incentives for researchers (Kuehne and Olden 2015).

The second method for providing PLSs in public libraries is making a workforce. This solution is based on a group of librarians or subject specialists formed within the library, who take responsibility for summarizing scientific findings in a way that is understandable to the general public. This specialized team should be formed in a way that can identify and summarize the scientific works needed by the community of users based on their demands. The advantage of forming such a team is that expertise is concentrated in the task of preparing high-quality PLSs, and after a while, it can be expected to publish proper PLSs. Therefore, training this team is easier than training various authors/researchers. Naturally, this team should include at least one professional researcher, a linguist, a subject specialist, and a science communicator. The summaries prepared by this team should be systematically evaluated and assessed. Perhaps sending PLSs to researchers who have authored scientific articles will be effective. These researchers can provide valuable feedback to the summarization team regarding the quality of a summary. Although this method involves providing proper PLSs based on the needs of the library and users, the costs of making this workforce and providing this service in a public library are likely to be higher than expected.

Another way to prepare PLSs is to involve library users. This method, which is closer to the concept of crowdsourcing, focuses on the assistance of library users in providing library services. In this method, library users who may have expertise in specific fields such as agriculture, physics, engineering, or other fields are asked to prepare PLSs for scientific works. However, it should be noted that not all library users might be familiar with plain language summary and its writing style, so it is better to provide specialized training to users interested in writing PLSs to improve their performance in preparing these summaries. After preparing a plain language summary for a specific work, it is better to have other users who have expertise in the same scientific field or in other areas such as writing, editing, or linguistics review the summaries to ensure their quality. One of the most significant advantages of this method is that it reduces the library's expenses in writing and preparing PLSs. Furthermore, this method allows for the utilization of various users' capabilities and encourages user engagement in facilitating library services.

However, it should be noted that not all users might be willing to engage in such tasks, so this method may not lead to a systematic and effective provision of PLSs in the library. Additionally, there is a concern that the PLSs may not be uniform in style since different users have different writing traditions, potentially resulting in a lack of consistency in the PLSs.

Another method to provide PLSs as a systematic service in public libraries is taking advantage of one of the Lay Text Summarization tools available (Vinzelberg et al. 2023). These tools utilize natural language processing algorithms to automatically generate PLSs from scientific articles. Public libraries can adopt and integrate these tools into their systems to provide quick and efficient PLSs to their users. This method eliminates the need for manual summarization and reduces the costs associated with training and maintaining a specialized workforce.

However, it is essential to consider the limitations of automated summarization tools. These tools may not always capture the nuanced meaning of the original scientific article accurately, and the generated PLSs may lack the human touch and clarity that a manual summary can provide. Therefore, it is crucial to evaluate the accuracy and comprehensibility of the PLSs produced by these tools before making them available to library users.

Finally, the last but not least method for providing PLSs involves collecting the existing accessible summaries that are currently available and published by academic journals. An increasing number of journal publishers are embracing the use of PLSs as a means to effectively communicate scientific discoveries to a wider audience. In this method, one or more librarians from a public library monitor and explore databases and academic journals, saving the published PLSs in a local repository. Then, using a certain mechanism, these summaries are made accessible to library users. The main advantage of this approach is that it reduces the library's expenses. Additionally, since academic journals continuously publish PLSs, the systematic provision of this service in the library can be ensured. However, an important concern in this regard is that the library cannot have a selection in the process of choosing works. In other words, the library must collect what is available rather than necessarily meeting the needs of the library and its users.

In general, the library has the option to choose one or more of these methods for providing PLSs. When selecting these methods, three key criteria should be considered: the accuracy and quality of the PLSs, the specific needs of the library and its users, and the associated costs of preparing and providing the PLSs. For instance, if the library is constrained by limited funds and does not anticipate the systematic provision of PLSs as a service, it may choose the first and third methods. This would involve seeking assistance from researchers who have authored scientific works or engaging expert library users in specific fields. However, if the library aims to establish a systematic and organized science promotion service, the second and fifth methods may be more suitable.

These methods entail establishing a dedicated team within the library to produce PLSs or leveraging existing PLSs through monitoring and exploration.

**Dissemination of PLSs through Public Libraries**

An important issue in developing the service of providing PLSs is how to disseminate these summaries and make them accessible to the users. Since many users of public libraries do not have access to specialized databases (such as, Web of Science, Scopus, PubMed, etc.) or reputable scientific references, or lack the knowledge to use them, disseminating PLSs through more general media becomes a significant issue that should receive attention in these libraries. Therefore, dissemination of PLSs through various channels should be considered in public libraries. Some of the most important of these channels are described below.

1) Dissemination through a dedicated (electronic/print/hybrid) magazine for PLSs: Perhaps the best option for disseminating PLSs in public libraries is preparing and publishing a dedicated magazine for this purpose. This magazine can be published monthly or at longer intervals, covering multiple PLSs in each issue. PLSs can be categorized and organized based on topics, users, or professional fields. However, it should be noted that launching this magazine will be effective when there are a significant number of PLSs in the library for publication in each issue; otherwise, other channels of dissemination will be more preferable. The most significant challenge of this method is the high costs associated with designing, publishing, and distributing the magazine.

2) Dissemination through library newsletters: Newsletters are one of the key publications prevalent in various types of libraries and are considered a marketing tool in these institutions. Public libraries that regularly publish newsletters can include PLSs within these media. Since newsletters often reach a wide range of users and library patrons are familiar with them, they provide a suitable platform for dissemination PLSs.

3) Dissemination through a dedicated online platform: If the library is looking for a cost-effective way to publish PLSs that provides easier access for users, taking advantage of internet facilities and online tools can be a suitable solution. In this regard, the library needs to create an online database and organize and publish PLSs on this platform. However, this method carries the risk that offline users may lack the knowledge, skills, or means to access this online platform and may not benefit from this service.

4) Dissemination through brochures: Another method for disseminating PLSs is to include these summaries in brochures created and published by the library. For example, a public library that aims to provide specific services or information to farmers can include one or more PLSs in a brochure designed for this purpose. However, it should be noted that these brochures are often designed thematically, and the specific plain language summary related to that theme should be included in them.

5) Dissemination through social media platforms: Nowadays, social media platforms have become efficient and effective tools for raising awareness about information resources and library services, as they have a wide user community and serve as an unparalleled means of accessing the target audience. Therefore, public libraries can share the prepared PLSs through these channels, which offer rapid, easy, and cost-effective dissemination.

In addition to the advantages of quick and easy dissemination through these networks, sharing and exchanging these summaries among diverse users is another benefit of this method.

6) In addition to the methods mentioned above for disseminating PLSs in public libraries, other approaches can also support these libraries in this endeavor. For example, disseminating PLSs in the form of posters within the library premises or public spaces can be another channel of dissemination. Some libraries may prefer to disseminate PLSs in audio form or as podcasts. In any case, public libraries should examine which channels are most suitable for their users and enhance the effectiveness of PLSs. Perhaps asking library users directly can provide librarians and library managers with more proper or innovative channels.

**Supporting Plain Language Summary Readers**

The service of providing PLSs in public libraries does not finish with the preparation and dissemination of these summaries. Since PLSs are usually very concise abstracts of scientific research, readers of these summaries may need more information to take more advantage of scientific research in their personal or professional lives. Therefore, public libraries should also consider provisions for supporting readers of PLSs.

The first issue in supporting readers of PLSs is answering to questions posed by these users. Some of these questions may be general (for example, "How can I access the full text of the research?" or "Can I do the experiments mentioned in the research on my own?") or and some of them may be more specific (for example, "How can I use the proposed pesticide in the research to eliminate agricultural pests?" or "The symptoms of my illness are similar to those mentioned in this research, can I use the recommended medication?"). In both cases, the library must have a certain procedure for answering these questions.

The second issue is receiving feedback, opinions, and suggestions that these readers have and want to share with the library. Some of this feedback or suggestions may relate to the research itself, while others may relate to the method of preparation and dissemination of these summaries. Feedback related to the research can be made available to researchers, journals, or publishers through a certain procedure, naturally leading to the long-term improvement in the quality of academic research.

Feedback related to the method of preparation and dissemination of PLSs can directly affect the quality of this service in the library, and can make the services of the library have a greater impact on the community. The library can go even further by involving library users and readers of PLSs in evaluating these writings to improve the quality of this service over time. Surveys, case studies, focus groups, and interviews with readers in this regard will be helpful for public libraries to improve the service of providing PLSs.

**Ethical and Legal Considerations**

It is necessary to acknowledge that the dissemination of PLSs in public libraries may come with challenges. The most significant challenge that may arise in this regard is that some users may practically use research findings without having a complete understanding of all aspects of those findings. This issue can pose risks to them. Additionally, it is not easy to apply some research findings in practice. Researchers are well aware of this issue and consider it when they take advantage of research findings. However, some users of public libraries who do not have complete knowledge in this field may overlook these considerations. For example, the recommended use of a specific medication or consumption of a particular food substance may be mentioned in a research, and public library users may follow that recommendation without paying attention to the potential risks. Therefore, librarians or personnel in charge of this service in public libraries must address this issue and provide the necessary awareness to users. Librarians and staff should constantly be in communication with users and readers of PLSs and the authors of university research papers to ensure that research findings are effectively and correctly utilized. Thus, the library should be able to establish a communication bridge between users and the researchers. This interaction will be beneficial for users and will also benefit researchers whose works are summarized. Researchers can obtain valuable feedback from their target audience and engage them more in academic research.

Additionally, copyright and intellectual property rights must be respected when preparing and disseminating these summaries. Public libraries should obtain proper permissions and licenses from the authors or publishers before summarizing and distributing PLSs. This not only ensures legal compliance but also acknowledges the intellectual contributions of the original researchers.

Obtaining permission from authors or publishers involves seeking their consent to summarize and (or) distribute PLSs of their works. This can be done through formal agreements or by following established guidelines provided by copyright laws and fair use provisions. Public libraries should make efforts to contact the relevant rights holders and negotiate appropriate terms for the use of copyrighted material. Moreover, public libraries should stay updated with changes in copyright laws and regulations to ensure compliance. They can seek legal advice or consult relevant copyright organizations to understand the specific requirements and obligations related to preparing and disseminating PLSs.

**PLS Service in Practice and Potential Outcomes**

In practice, implementing PLSs within library collections, systems, and discovery requires coordination between various stakeholders, including librarians, metadata specialists, cataloging staff, and potentially researchers and publishers. Libraries need to establish guidelines and standards for creating and integrating PLSs into their existing workflows, metadata schemas, and discovery interfaces. Additionally, libraries should consider evaluating the impact and effectiveness of PLSs through user feedback, usage statistics, and ongoing assessment to ensure continuous improvement and alignment with user needs.

Public library practitioners can apply the proposed framework for providing PLSs in public libraries by through several steps. First, library practitioners should familiarize themselves with the PLS concept. Library practitioners should understand the importance and benefits of PLSs in bridging the gap between scientific research and the public. They should be aware of the role public libraries can play in promoting science and making research findings more accessible to library users. Second, library practitioners should explore the various methods discussed in the paper for providing PLSs in public libraries. Practitioners should assess the feasibility and suitability of each method based on their library's resources, goals, and user needs. Consequently, based on their assessment, library practitioners should select one or more methods that align with their library's objectives and capabilities. They should consider factors such as the accuracy and quality of the summaries, the needs of their users, and the associated costs. The selected method(s) should be practical and feasible to implement within the library's context. Develop necessary infrastructure: Once the method(s) are chosen, library practitioners need to develop the necessary infrastructure to support the provision of PLSs. This may involve establishing partnerships with researchers and authors, creating a dedicated team within the library, providing training for authors and users, integrating automated summarization tools into library systems, or setting up mechanisms to collect and disseminate existing summaries.

Furthermore, library practitioners should collaborate with researchers, authors, and library users to ensure the successful implementation of the chosen method(s). They should establish communication channels with researchers to request PLSs for relevant research findings. If a specialized team is formed, practitioners should ensure the team members have the required expertise and support their ongoing professional development. Involving library users in the process can be done through training programs, crowdsourcing efforts, or seeking user feedback on the summaries. In addition, library practitioners should develop strategies to disseminate the PLSs effectively. This may involve creating a dedicated magazine or newsletter for PLSs, organizing them by topics or fields of interest, and making them available in both print and electronic formats. Practitioners should leverage existing library channels such as newsletters, brochures, social media, library websites, and online databases to make the summaries accessible to library users.

Last but not least, Library practitioners should continuously evaluate the effectiveness and impact of providing PLSs in their library. They can collect user feedback, conduct surveys or interviews, and analyze usage statistics to assess the usefulness of the summaries and identify areas for improvement. Regularly reviewing and updating the framework based on user needs and emerging trends will help ensure the ongoing success of the service.

Probably, providing PLSs in public libraries brings several potential outcomes of the user experience. These outcomes can provide insights into the potential impact and benefits of implementing such initiatives. One of the primary goals of providing PLSs is to enhance user understanding of complex research findings. Similar studies have shown that when users have access to PLSs, they are more likely to comprehend and engage with the information (Dormer et al. 2022). By presenting research in a clear and accessible manner, public libraries can contribute to users' understanding of scientific concepts and findings. In addition, PLSs aim to make scientific research more accessible to a broader audience, including individuals with limited scientific or technical background knowledge. Studies have indicated that providing PLSs can help bridge the gap between academic research and the general public, ensuring that important findings are not confined to specialized communities (Stoll et al. 2022). This increased accessibility can empower public library users to explore and benefit from a wider range of research. Improving accessibility to research findings will increase the visibility of public libraries in their target communities.

Access to PLSs can enable public library users to make more informed decisions. By presenting research findings in a user-friendly format, libraries can empower individuals to evaluate the relevance and implications of scientific research for their personal or professional lives. Users can use PLSs to make informed choices, such as adopting evidence-based practices or making decisions related to health, technology, or economy. Furthermore, studies have shown that providing PLSs can promote greater engagement and interest in scientific research among public (Kuehn 2017). By making research more approachable, public libraries can encourage users to explore a wider range of topics, develop a curiosity for scientific knowledge, and potentially foster a culture of lifelong learning. This engagement with research can have long-term benefits for individuals and society as a whole. PLSs can empower public library users to actively participate in discussions related to scientific research. When individuals have access to clear and concise summaries, they are better equipped to contribute to public debates, engage in citizen science initiatives, or participate in community-driven decision-making processes. Libraries can facilitate informed participation by providing PLSs that enable users to understand and contribute to scientific discourse. Considering the mission of public libraries to inspire lifelong learning and strengthen their users and communities in all required areas, it is important that these initiatives bring more new and accurate knowledge to their patrons.

Furthermore, "*PLS […] address information equity (along with other key drivers such as openness and discoverability) by helping to bridge the information gap between specialist professionals and other stakeholders, regardless of background or level of experience*." (Rosenberg et al. 2023). Therefore, by providing PLSs, public libraries can contribute to bridging the information gap that often exists between specialist professionals and other stakeholders. Public libraries are known for their commitment to providing access to information for all members of the community (International Federation of Library Associations and Institutions (IFLA) 2015). By offering PLSs, libraries can ensure that research findings and scholarly knowledge are not limited to a select group of specialists but are made accessible to a broader audience.

**Conclusion**

The global spread of the "COVID-19" virus, which has been ongoing for over three years, is just one of the events that prove the dependence of societies - especially the general public, who have relatively limited access to accurate sources of knowledge - on science and research findings. During this pandemic period, many have looked at universities, research institutes, and educational and research centers and expected researcher to find solution to remove this virus or at least mitigate the impact and consequences of this global crisis. Recent research shows that people monitor scientific findings and may apply them in their daily lives. For example, the study by Obiała, et al. (2021) demonstrates the attention of the public to the science and shows that online social networks have become a key platform for the general public's access to research findings.

Public libraries have been referred to as universities of the public. Therefore, libraries should be the primary reference and institution that enables public access to the latest scientific and research findings. Many users of public libraries rely solely on these institutions to fulfill their information needs, especially those who have limited technological literacy. Thus, public libraries should provide a space where these users can access the latest scientific findings that are relevant to their lives and professions. It seems that the publication of PLSs can be an important and valuable solution for public access to scientific findings, and public libraries can play a key role in promoting science.

Providing PLSs as a new service by public libraries is noteworthy from another perspective as well. Since these summaries can increase the impact of research, it can be expected that research will have a greater influence on society through the publication of PLSs. In other words, this service can bring science and research findings closer to the societies and the general public, making research results directly relevant to people's daily lives. Taking this approach to PLSs can be helpful in marketing programs for this service.

However, there are some legal and ethical considerations that public libraries should take into account when providing PLSs. Challenges such as the accuracy and reliability of the information presented in the summaries, copyright and intellectual property right, privacy and data protection and transparency about the limitations of PLSs. While these summaries aim to make research more accessible, they may not capture the full complexity or nuances of scientific studies. It is important to educate library users about the purpose and scope of these summaries, encouraging them to explore the original research for a more comprehensive understanding.

Public libraries have several methods to provide PLSs of scientific works. These include asking author and researchers themselves, forming a specialized workforce within the library, involving library users in the summarization process, utilizing automated lay text summarization tools, or collecting existing PLSs. Each method has its advantages and considerations in terms of cost, quality, and user engagement. Public libraries should carefully assess their resources, user preferences, and goals to determine the most suitable approach for providing PLSs as a valuable service to their communities.

It should be noted that providing PLSs in public libraries require further research to highlight various aspects of this service for librarians and public library managers. Perhaps, periodic evaluations should be conducted after the implementation of these services in order to further development. Probably, providing PLSs in public libraries can transform these libraries into a more accelerated gateway to public knowledge and an effective tool for increasing the impact of research in society.

Wynia Baluk, Kaitlin, Nicole K. Dalmer, Leora Sas van der Linden, Lisa Radha Weaver, and James Gillett. 2023. "Towards a research platform: partnering for sustainable and impactful research in public libraries." *Public Library Quarterly* 42 (1): 71-91. https://doi.org/10.1080/01616846.2022.2059315.

https://doi.org/10.1080/01616846.2022.2059315.

**Appendix 1**

Table 1. A list of guidelines for preparing a lay summery.

| Link to Plain language summary Preparing Guidelines (Last visit: June 9, 2023) |
|---|
| https://training.cochrane.org/handbook/current/chapter-iii-s2-supplementary-material |
| https://www.agu.org/Share-and-Advocate/Share/Community/Plain-language-summary |
| https://authorservices.taylorandfrancis.com/publishing-your-research/writing-your-paper/how-to-write-a-plain-language-summary/ |
| https://cdnsciencepub.com/authors-and-reviewers/writing-a-plain-language-summary |
| https://perspectivesblog.sagepub.com/blog/author-services/how-to-write-a-plain-language-summary-for-journal-articles |
| https://www.nihr.ac.uk/documents/plain-english-summaries/27363 |
| https://www.plainlanguagesummaries.com/plain-language-summaries-how-to-write/ |
| https://www.hra.nhs.uk/planning-and-improving-research/best-practice/writing-plain-language-lay-summary-your-research-findings/ |
| https://www.cochrane.no/how-write-plain-language-summary |
| https://elifesciences.org/articles/25410 |
| https://courses.karger.com/courses/how-to-write-a-plain-language-summary |
| https://www.jrheum.org/content/plain-language-summaries |

| Link to Plain language summary Preparing Guidelines (Last visit: June 9, 2023) |
|---|
| https://www.cfn-nce.ca/wp-content/uploads/2017/09/cfn-guidelines-for-lay-summaries.pdf |
| https://www.britishecologicalsociety.org/wp-content/uploads/2020/07/Plain-Language-Summary-Guidelines_2020.pdf |
| https://learning.edanz.com/plain-language-summary-abstract/ |
| https://cimvhr.ca/forum/resources/WritingTheLaySummaryBasics.pdf |
| https://involvementtoolkit.clinicaltrialsalliance.org.au/toolkit/undertaking/writing-in-plain-language/ |
| https://www.ctontario.ca/patients-public/resources-for-engaging-patients/toolkit-to-improve-clinical-trial-participants-experiences/plain-language-summaries/ |
| https://www.hunter.cuny.edu/rwc/handouts/the-writing-process-1/invention/Guidelines-for-Writing-a-Summary |
| https://intelligentediting.com/blog/8-hacks-to-nail-your-next-plain-language-summary/ |
| https://www.gov.uk/government/publications/accessing-ukhsa-protected-data/approval-standards-and-guidelines-lay-summary |
| https://www.kellogg.edu/upload/eng151/chapter/how-to-write-a-summary/index.html |
| https://europepmc.org/docs/A2U_2014_plain_English_writing_guidance.pdf |
| https://communications.admin.ox.ac.uk/resources/how-to-guide-writing-a-lay-summary |

| Link to Plain language summary Preparing Guidelines (Last visit: June 9, 2023) |
|---|
| https://honorscollege.charlotte.edu/sites/honorscollege.charlotte.edu/files/media/Honors%20application%20Layman%20Summary.pdf |
| https://www.icgc-argo.org/page/141/e86-lay-summary-guide-for-researchers-tips-on-writing-a-lay-summary- |
| https://www.certara.com/blog/writing-plain-language-summaries-clinical-study-results/ |
| https://www.isoqol.org/plain-english-summary-writing-tips/ |
| https://kidney.ca/getattachment/Research/Funding-Opportunities/Accordion/Allied-Health-Kidney-Scholarships/Best-Practices-FINAL-English.pdf?lang=en-CA |
| https://www.exeter.ac.uk/media/universityofexeter/research/microsites/treetranslationalresearchexchangeexeter/Writing_a_Lay_Summary.pdf |
| https://www.dcc.ac.uk/guidance/how-guides/write-lay-summary |
| https://www.psycharchives.org/en/item/a315b937-dc95-43ae-ad72-c5c707443c64 |
| https://arthropodecology.com/2013/08/01/a-guide-for-writing-plain-language-summaries-of-research-papers/ |
| https://www.diabetes.org.uk/research/for-researchers/apply-for-a-grant/general-guidelines-for-grant-applicants/tips-on-writing-a-lay-summary |
| https://jmvfh.utpjournals.press/jmvfh/resources/lay-summary |
| https://km4s.ca/wp-content/uploads/E4D-Plain-Language-Summaries-Toolkit-General.pdf |
| https://www.srcd.org/public-summaries-writing-lay-audience |

| Link to Plain language summary Preparing Guidelines (Last visit: June 9, 2023) |
|---|
| https://www.enago.com/academy/great-lay-summary-top-09-tips-to-write/ |
| https://www.elsevier.com/connect/authors-update/in-a-nutshell-how-to-write-a-lay-summary |
| https://sixdegreesmed.com/six-tips-for-developing-a-plain-language-summary/ |
| https://www.editage.com/blog/how-to-write-a-research-paper-summary/ |
| https://fapp.ucsd.edu/HowToWrite-PLS-2019-digital-final.pdf |
| https://www.campbellcollaboration.org/images/How_to_write_a_Campbell_PLS.pdf |